\documentclass[PRL, reprint,10pt,aps,
, amsmath,amssymb, longbibliography]{revtex4-1}

\usepackage{graphicx}
\usepackage{dcolumn}
\usepackage{bm}

\newcommand{\tadv}{\theta_\text{adv}}
\newcommand{\trec}{\theta_\text{rec}}
\newcommand{\tadvrec}{\theta_\text{adv, rec}}
\newcommand{\dcos}{\Delta \cos \theta}

\usepackage[utf8]{inputenc}
\usepackage[english]{babel}

\usepackage{xcolor, soul}
\sethlcolor{white}

\begin{document}


\title{Mapping micron-scale wetting properties of superhydrophobic surfaces}

\author{Dan Daniel$^{1}$}
 \email{daniel@imre.a-star.edu.sg}
\author{Chee Leng Lay$^{1}$}
\author{Anqi Sng$^{1}$}
\author{Corryl Jing Jun Lee$^{1}$}
\author{Darren Chi Jin Neo$^{1}$}
\author{Xing Yi Ling$^{2}$}
\author{Nikodem Tomczak$^{1}$}
  
\affiliation{$^{1}$Institute of Materials Research and Engineering, A*STAR (Agency for Science, Technology and Research), 2 Fusionopolis Way, Innovis, Singapore 138634}
\affiliation{$^{2}$Division of Chemistry and Biological Chemistry, School of Physical and Mathematical Sciences,
Nanyang Technological University, 21 Nanyang Link, Singapore 637371}

\begin{abstract}
There is a huge interest in developing super-repellent surfaces for anti-fouling and heat-transfer applications. To characterize the wetting properties of such surfaces, the most common approach is to place a \textit{millimetric}-sized droplet and measure its contact angles. The adhesion and friction forces can then be inferred indirectly using the Furmidge's relation. While easy to implement, contact angle measurements are semi-quantitative and cannot resolve wetting variations on a surface. Here, we attach a \textit{micrometric}-sized droplet to an Atomic Force Microscope cantilever to directly measure adhesion and friction forces with nanonewton force resolutions. We spatially map the micron-scale wetting properties of superhydrophobic surfaces and observe the time-resolved pinning-depinning dynamics as a droplet detaches from or moves across the surface. 
\end{abstract}
\maketitle

\section{Introduction}

Water droplets can easily bounce and roll off superhydrophobic surfaces, enabling many important applications, ranging from water-repellent coatings to drag-reduction in ships \cite{quere2008wetting, bocquet2011smooth}. Since Thomas Young first proposed the concept in his seminal 1805 paper \cite{young1805}, contact angle measurements have become the gold standard to characterize surface wetting property. In fact, superhydrophobic surfaces are often defined by their high contact angles $\theta > 150^{\circ}$ \cite{quere2008wetting, reyssat2010dynamical, de2004capillarity}.

There are a few reasons for the popularity of contact angle measurements. Firstly, they are relatively easy to implement, requiring only good lighting and a high-resolution camera. Moreover, the adhesion and friction forces can be inferred, albeit indirectly, from the advancing and receding contact angles $\tadvrec$  \cite{samuel2011study, furmidge1962studies}.


There is, however, a growing debate as to whether contact angles adequately describe the surface wetting properties \cite{decker1999physics, schellenberger2016water, daniel2018origins}. They do not, for example, capture the local wetting variations due to chemical heterogeneity or surface texture \cite{liimatainen2017mapping}. Moreover, for large contact angles (e.g. in superhydrophobic surfaces), a small error in the positioning of the droplet base (as small as a single pixel) translates to  a large error in the contact angle value (of more than $10^{\circ}$) \cite{Liu1147, srinivasan2011assessing}. 

To overcome these limitations, there have been several attempts to develop more sensitive surface characterization techniques. The most common approach is to use a cantilever force sensor to directly measure the friction and adhesion forces (with a typical 0.1 $\mu$N resolution) acting on a \textit{millimetric} droplet \cite{pilat2012dynamic, t2013electrically, daniel2017oleoplaning}. Recently, by greatly suppressing the environmental noise, our group has improved the resolution of such an instrument (which we named the Droplet Force Apparatus) to about 5 nN \cite{daniel2019hydration}. More impressively, by combining a sensitive force sensor with a motorized sample stage, Ras and co-workers were not only able to achieve a similar nN force resolution, but were also able to map wetting variations with a lateral resolution of $10$ $\mu$m \cite{liimatainen2017mapping}. 

Since its invention in 1986, Atomic Force Microscopy (AFM) has become a standard and powerful surface characterization tool \cite{binnig1986atomic}. AFM was used to characterize surface wetting properties by measuring the tip-surface interactions \cite{rana2016correlation, dupres2007wetting, eastman1996adhesion, delmas2011contact}; however, the AFM tip is typically made of solid silicon/silicon nitride and has a pyramidal geometry, which poorly approximates droplet-surface interactions.

In this paper, we replace the solid tip with a microdroplet and use this modified droplet probe AFM technique to map local wetting properties of superhydrophobic surfaces \cite{xie2017surface, shi2016long, manor2008hydrodynamic}. By attaching a 40 wt$\%$ glycerin-water droplet of diameter 20--50 $\mu$m onto the AFM cantilever, we are able to measure the friction and adhesion forces with much improved force and lateral resolutions (at least 1 nN and 1 $\mu$m, respectively), and observe the fast pinning-depinning dynamics (millisecond timescale) as a droplet detaches from or moves across the surface.



\section{Quantifying local wetting properties}

\begin{figure*}[!htb]
\centering
\includegraphics[]{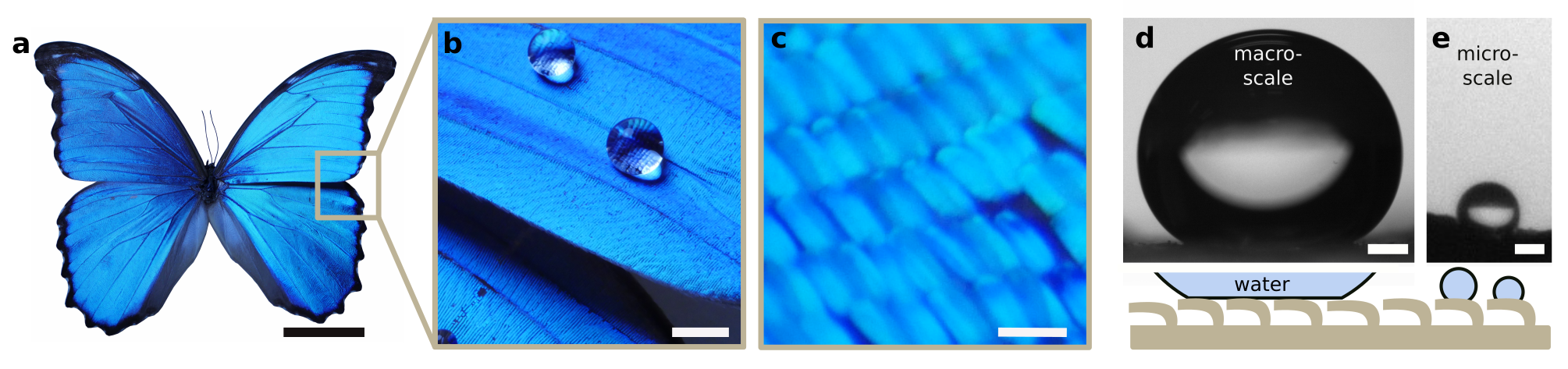}
\caption{(a, b) Millimetric-sized water droplets bead up on the wings of a morpho butterfly, which are covered by (c) superhydrophobic scales. Scale bars are 4 cm, 3 mm, and 150 $\mu$m, respectively. (d) The scales of the butterfly wing trap an air layer, resulting in a high contact angle of 160 $\pm$ 10$^{\circ}$ for millimetric droplets. Scale bar is 0.5 mm. (d) On individual scale, it is difficult to measure the contact angles of a micrometric droplet. Scale bar is 50 $\mu$m }
\label{fig:morpho_scheme}
\end{figure*}

We first investigated the wetting properties of the Morpho butterfly wings, well-known for their brilliant blue colour and excellent water repellency (Fig.~\ref{fig:morpho_scheme}a--c) \cite{lei2006morpho, niu2015excellent}. The wings are covered with scales with intricate micro- and nano-structures (See Supplementary Figure \hl{S1} for detailed scanning electron micrographs), which trap a stable air layer, resulting in superhydrophobicity and a high static contact angle of 160 $\pm$ 10$^{\circ}$  for millimetric water droplets (Fig.~\ref{fig:morpho_scheme}d). 

Conventionally, the forces required to remove the droplets are inferred \textit{indirectly} from the advancing and receding contact angles $\tadvrec$ \cite{huhtamaki2018surface, Liu1147, srinivasan2011assessing}. For example, the friction force $F_{\text{fric}}$ required to move the droplet laterally is given by the Furmidge's relation
\begin{equation} \label{eq:Furmidge}
F_{\text{fric}} = \pi \gamma r \dcos = \pi \gamma r (\cos \trec - \cos \tadv), 
\end{equation}
where $r$ is the droplet's base radius, $\gamma$ is the surface tension, and $\dcos = \cos \trec - \cos \tadv$ is the contact angle hysteresis \cite{butt2014characterization, samuel2011study, furmidge1962studies}. Contact angles measurements can be performed relatively easily for millimetric droplets, but become very challenging for micrometric droplets (Supplementary Figs.~\hl{S2, 3}). Moreover, the droplet's base can be obscured on uneven surfaces, further complicating contact angle measurements (Fig.~\ref{fig:morpho_scheme}e).
 
To overcome the limitations outlined above, we propose using droplet probe AFM to \textit{directly} measure adhesion and friction forces $F_{\text{adh, fric}}$ by monitoring the flexular and torsional deflections of an AFM cantilever, respectively. This is done by shining a laser light (infra-red, wavelength of 980 nm) onto the cantilever, which is reflected into a four-quadrant sensor. To convert the raw voltage signals $V_{\text{vert, lat}}$ into forces $F_{\text{adh, fric}}$, we use the Sader's method \cite{sader1998frequency, green2002torsional, wagner2011noncontact, butt2005force}. See Supplementary Figures~\hl{S4, 5} for details.

Figure \ref{fig:morpho_adhesion}a is a schematic of how $F_{\text{adh}}$ is measured. We first attached a 40 wt$\%$ glycerin-water droplet of diameters 20--50 $\mu$m onto a tipless cantilever probe with a flexular spring constant of $k_z = 2$ N/m. The addition of glycerol suppresses the evaporation rate of the microdroplet, without greatly changing its surface tension or viscosity (69 mN/m and 4 mPa.s, compared to 73 mN/m and 1 mPa.s for pure water). From the flexular deflection of the cantilever $\Delta z$, we can then deduce the force on the droplet $F = k_z \Delta z$ as it approaches and retracts from the surface at a controlled speed $U$ = 1--50 $\mu$m/s. Since the microdroplet is smaller than the size of the wing-scale, we are able the quantify local wetting properties of individual scales  (Fig.~\ref{fig:morpho_adhesion}b).   
 
Figure \ref{fig:morpho_adhesion}c shows the force spectroscopy results for a 30-$\mu$m-diameter droplet with volume $V$ = 20 $\pm$ 2 pl approaching and retracting from one of the wing-scales at $U$ = 10 $\mu$m/s. The droplet's size and volume can be obtained by optical microscopy and using Cleveland's method (Supplementary Fig.~\hl{S6}) \cite{butt2005force}. When the droplet is far from the surface, the AFM did not detect any force, i.e. $F$ = 0; however, upon contact, there is a sudden attractive snap-in force $F_{\text{snap}}$ = 132 nN. We continue to press onto the microdroplet to the maximum normal force $F_{\text{N}}$ = 10 nN, before retracting (solid line). For the droplet to be completely detached from the surface, there is a maximum adhesion force that must be overcome $F_{\text{adh}}$ = 720 nN. By integrating under retract curve, we can also obtain the amount of work required to remove the droplet $W_{\text{adh}}$ = 5.6 pJ, which is a fraction of the total surface energy of the droplet $4 \pi R^{2} \gamma \approx$ 200 pJ, reflecting the liquid-repellent nature of the surface.  

\begin{figure*}[!htb]
\centering
\includegraphics[]{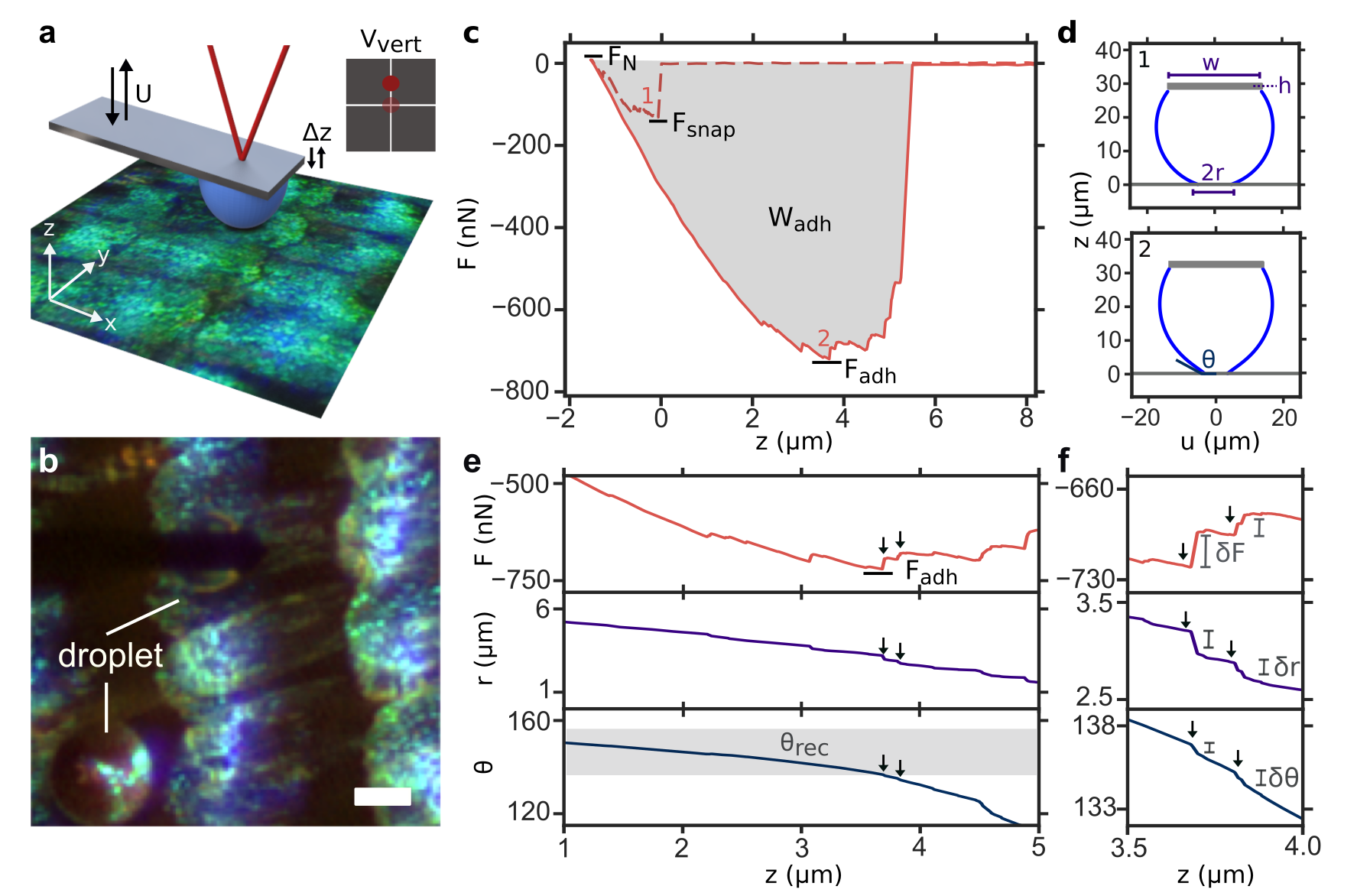}
\caption{(a) The adhesion force $F_{\text{adh}}$ on individual scales of the butterfly wing can be measured accurately with droplet probe AFM. (b) Top-down view of the setup, showing one droplet attached to an AFM cantilever and another sitting on a wing-scale. Scale bar is 50 $\mu$m. (c) Force spectroscopy of a 30 $\mu$m-sized droplet on a wing-scale. Dashed and full lines are the approach and retract curves. The droplet position $z$ has been corrected for cantilever deflection. By solving the Young-Laplace equation, we can deduce (d) the droplet geometry, and (e, f) obtain the the base radius $r$ and the contact angle $\theta$ at different time points.}
\label{fig:morpho_adhesion}
\end{figure*}
\begin{figure*}[!htb]
\centering
\includegraphics[]{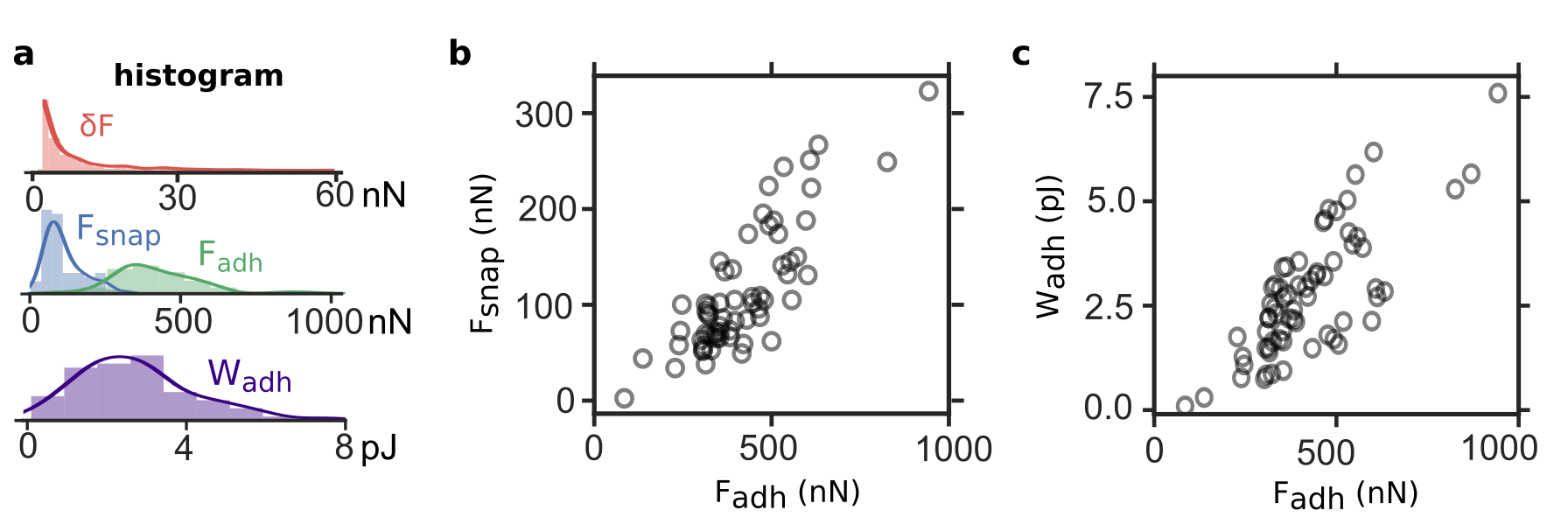}
\caption{(a) Histogram of $\delta F$, $F_{\text{snap, adh}}$, and $W_{\text{adh}}$ for a total of 66 different scales and 670 depinning events. (b, c) Plots of $F_{\text{adh}}$ against the snap-in force $F_{\text{snap}}$ and $W_{\text{adh}}$, with each point representing different scale.}
\label{fig:morpho_stat}
\end{figure*}

It is also possible to relate the force spectroscopy measurements with the evolution of the microdroplet's contact angle with time. The shape of the droplet $u(z)$ is described by the axisymmetric Young-Laplace equation  
\begin{equation} \label{eq:laplace}
\begin{split}
\frac{u''}{(1+u'^{2})^{3/2}} - \frac{1}{u\sqrt{1 + u'^{2}}} &= -\Delta P/\gamma \; \mathrm{for} \; z \in (0, h),\\
	u(h) &= a \\
	u(0) &= r \\
	\int_{0}^{h} \pi u^{2}dz &= V \\ 
	F = -2 \pi \gamma r \sin &\theta + \pi r^{2}\Delta P. 
\end{split}
\end{equation}
The droplet contact area on the cantilever is fixed by the cantilever's width $2a$ = 28 $\mu$m (Supplementary Fig.~\hl{S7}), while the droplet height $h$ and the force $F$ can be deduced from the position of the piezomotor and the cantilever deflection, respectively. By solving  equation \ref{eq:laplace} numerically, we can deduce the droplet's geometry, and in particular the Laplace pressure inside the droplet $\Delta P$, the base radius $r$, and the contact angle $\theta$ at different time points (Fig.~\ref{fig:morpho_adhesion}d).  

For example, during snap-in (time point 1), the droplet has a base radius of $r$ = 4.5 $\mu$m and contact angle of $\theta$ = 161.3$^{\circ}$, respectively; whereas at the maximum $F = F_{\text{adh}}$, $r$ = 3.2 $\mu$m and $\theta$ = 137.0$^{\circ}$. See Supplementary Figs.~\hl{S7} and \hl{S8} for details of the numerical method used, as well as \cite{Github} for the implementation in Python. Supplementary Movie 1 shows the full simulation of the droplet geometry during the force spectroscopy measurement.    

During the force spectroscopy measurements, the flexular deflections can be monitored with high speeds (up to hundreds of kHz). This allows us probe the fast pinning-depinning dynamics (and stick-slip motion) of the contact line with unprecedented details. As the droplet retracts, both $r$ and $\theta$ decreases gradually; except at certain time points (which correspond to discrete force jumps $\delta F \sim$ nN), where there are sudden discontinuities in $r$ and $\theta$ over a timescale $\Delta t \sim$ ms. The magnitudes of the discontinuities $\delta r, \delta \theta$ can be as small as fractions of a micron and a degree, respectively (Fig.~\ref{fig:morpho_adhesion}e, f). Note that $r$ and $\theta$ should be taken as effective, radially-averaged values, since we made the assumption of axisymmetric droplet's shape, when the contact line is likely to be jagged and discontinuous \cite{daniel2018origins}.

We also note that there is no single value of $\trec$ for the microdroplet. If we consider the retraction curve up to $F_{\text{adh}}$, where the contact line retracts at a relatively constant speed of $|\mathrm{d}r/\mathrm{d}t| \approx$ 0.7 $\mu$m/s, $\trec$ varies between 137$^{\circ}$ and $150^{\circ}$ (shaded area in Fig.~\ref{fig:morpho_adhesion}e), which translates to a contact angle hysteresis value of $\Delta \cos \approx 1 + \cos \trec$ = 0.13--0.27 \cite{schellenberger2016water}, which is slightly higher than the $\Delta \cos = 0.05$--0.21 measured for millimetric-sized droplet using conventional tilting-plate method (Supplementary Fig.~S2). This could be explained by the fact that the overlapping wing-scales are able to trap an additional air layer, resulting in lower solid surface fraction for millimetric-sized droplets.

We repeated the force spectroscopy measurements for a total of 66 different wing-scales (with 670 depinning events), whose results are summarized in Fig.~\ref{fig:morpho_stat}a. There are significant wetting variations between scales. For example, $F_{\text{snap}}$ can be as small as a few nN to more than 300 nN for the same 20 pl microdroplet; similarly $F_{\text{adh}}$ can vary between 84 and 943 nN, while $W_{\text{adh}}$ varies between 0.1 and 7.6 pJ. During droplet retraction, the magnitude of the force jumps $\delta F$ can vary between a couple of nN to more than 60 nN. Since $\delta F \sim a_{\text{p}} \gamma n_{\text{p}}$ where $a_{\text{p}}$ is the typical size and $n_{\text{p}}$ is the number of pinning points contributing to the depinning event, larger $\delta F$ is less likely to occur, because it involves simultaneous depinning from multiple points. Hence, the probability density of $\delta F$ decreases rapidly with incrasing $|\delta F|$.           


In general, higher $F_{\text{adh}}$ translates also to higher $F_{\text{snap}}$ and $W_{\text{adh}}$ values, though there are large variations (Fig.~\ref{fig:morpho_stat}b, c). For example, when $F_{\text{adh}}$ = 500 nN, $F_{\text{snap}}$ vary between 50 and 200 nN, while $W_{\text{adh}}$ vary between 1 and 5 pJ.           

The force spectroscopy measurements therefore provide us with a wealth of information that is not easily obtained using conventional contact angle measurements. Note that while the results vary between wing-scales, the force spectroscopy curves for each scale are reproducible (Supplementary Fig.~\hl{S9}). Experimentally, we also found that the results do not depend on the speed $U$, i.e. viscous effects are not important, or the applied $F_\text{N}$ (Supplementary Fig.~\hl{S10}).  

\section{Mapping Micron-scale wetting variations}

We can take advantage of the raster scanning capability of the AFM and perform force spectroscopy measurements over a grid array of points to map micron-scale wetting variations on surfaces. To illustrate this, we chose a superhydrophobic surface with well-defined micro-/nano-structures, which consists of a square array of 10 $\mu$m-diameter pillars decorated with smaller 2 $\mu$m pillars, spaced $L$ = 20 $\mu$m and $l$ = 3.5 $\mu$m apart, and with a total height $H$ = 2.5 $\mu$m (Fig.~\ref{fig:pillar_adhesion}a). See Supplementary Figure~\hl{S11} for scanning electron micrographs and AFM images. 

Figure \ref{fig:pillar_adhesion}b is an adhesion map for an array of 3 $\times$ 3 pillars (70 $\times$ 70 $\mu$m, 100 $\times$ 100 px). The force spectroscopy measurements were performed with a 30-$\mu$m-sized droplet and at a relatively high speed $U$ = 150 $\mu$m s$^{-1}$ to minimize the scanning time to about 20 minutes. The adhesion is greatest when the droplet is in the gap between four neighbouring pillars (marked 1 in Fig.~\ref{fig:pillar_adhesion}b), where $F_\text{adh}$ can reach almost 800 nN (Fig.~\ref{fig:pillar_adhesion}d). In contrast, when the droplet is on top of the micropillar (marked 2 in Fig.~\ref{fig:pillar_adhesion}b), $F_\text{adh}$ is only about 150 nN. 

It is likely that the 30-$\mu$m-sized droplet is touching the base at area 1, where the gap between the pillars is the largest. In fact, when the experiment is repeated on a micropillar surface with the exact same lateral dimensions but taller $H$ = 12 $\mu$m, there are no longer high adhesion areas in area 1 (Supplementary Fig.~\hl{S12}). A closer look at the adhesion map also reveals the four-fold and six-fold symmetries that reflects the underlying positional symmetries of the micropillars (Fig.~\ref{fig:pillar_adhesion}c).

We have therefore demonstrated the ability of droplet probe AFM to map wetting variations in the micron-scale, despite the droplet's size being much larger. This is because the lateral resolution is determined the droplet's base diameter $2r \sim$ 1 $\mu$m (rather than its outer diameter $2R$) just before it detaches (Supplementary Movie 1).   

\begin{figure*}[!htb]
\centering
\includegraphics[]{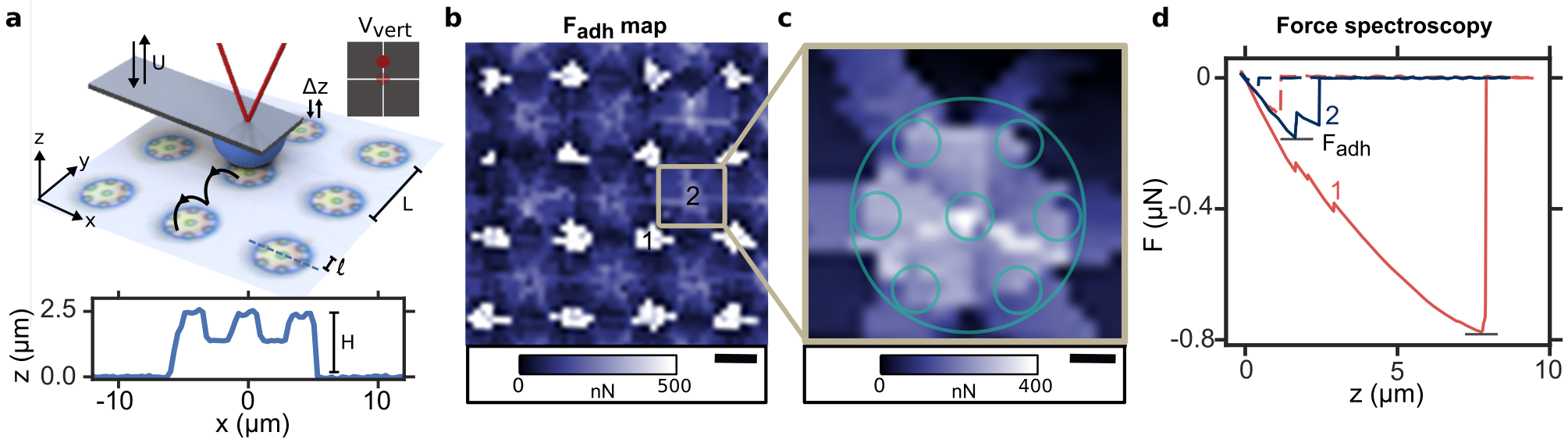}
\caption{(a) $F_{\text{adh}}$ of a micro-droplet on a structured surface can be mapped with micron-scale resolution. The surface consist of a square array of 10 $\mu$m-diameter pillars with smaller 2 $\mu$m pillars on them. (b) Force maps for 30 $\mu$m-sized droplet. Scale bar is 10 $\mu$m. (c) A more detailed force map with the outline of the micropillars superimposed. Scale bar is 2 $\mu$m. (d) Force spectroscopy curves on top (area 1) and in between pillars (area 2).}
\label{fig:pillar_adhesion}
\end{figure*}
\begin{figure*}[!htb]
\centering
\includegraphics[scale=1.1]{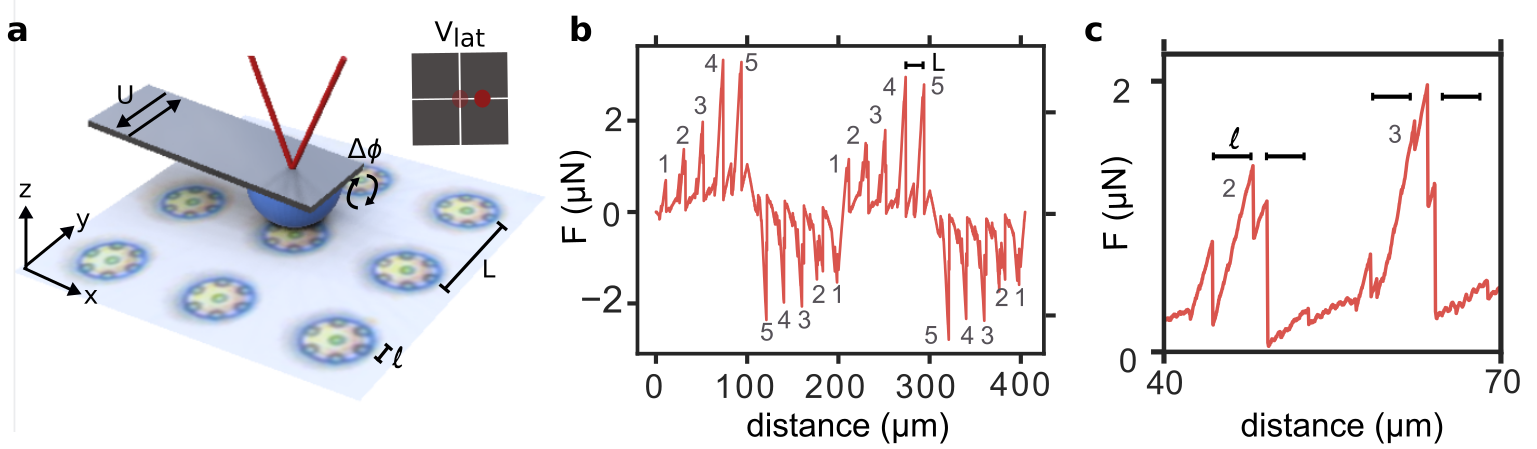}
\caption{(a) $F_{\text{fric}}$ of a micro-droplet moving on the micropillar surface can be measured accurately with droplet probe AFM. (b, c) Depinning due to individual 10 $\mu$m and 2 $\mu$m micropillars (spaced $L$ = 20 $\mu$m and $l$ = 3.5 $\mu$m apart, respectively) can be clearly distinguished. }
\label{fig:pillar_friction}
\end{figure*}

\section{Measuring friction forces}

We can measure the friction force $F_{\text{fric}}$ by moving the droplet laterally and monitoring the resultant torsional deflection of the cantilever $\Delta \phi$, since $F_{\text{fric}} = k_{\phi} \Delta \phi/h$, where $k_{\phi} = 69.3$ nN m is the torsional spring constant.

Figure \ref{fig:pillar_friction}b shows the lateral force measured for an array of five pillars (labelled 1--5), as we move a 30-$\mu$m-sized droplet back and forth over a 100 $\mu$m distance twice. During the motion of $U$ = 5 $\mu$m s$^{-1}$, the normal force is kept at $F_{\text{N}}$ = 5 nN. The force measured is positive one way and negative the other way, because the torsional deflections are in the opposite directions. Note also that since the spacing between pillars is $L$ = 20 $\mu$m, the droplet is in contact with only one pillar at any one time. 

The force required to detach from one pillar can vary between $F_{\text{fric}} = 1.3 \pm 0.3$ $\mu$N (pillar 1) and 2.7 $\pm$ 0.2 $\mu$N (pillar 5). A closer look at the force spectroscopy measurements also reveal force jumps spaced $l = 3.5$ $\mu$m apart, due to depinning from the smaller 2-$\mu$m-sized pillars (Fig.~\ref{fig:pillar_friction}c). Experimentally, we also found that, $F_{\text{fric}}$ is independent of $U$, consistent with previous reports (Supplementary Fig.~\hl{S13) }\cite{daniel2018origins}.       

\section{Discussion and conclusion}

To conclude, we are able to use droplet probe AFM to map micron-scale wetting properties of the superhydrophobic surfaces and also to probe the fast pinning-depinning dynamics during droplet detachment. In addition, we are able to relate the force spectroscopy measurements to conventional contact angle values by solving the Young-Laplace equation. While we have confined our discussion to superhydrophobic surfaces, the technique described in this paper can easily be adapted for other liquid probes (e.g. oil droplet) and other classes of liquid-repellent surfaces (e.g. underwater superleophobic surface). 

Droplet probe AFM technique described here will greatly complement other ultra-sensitive surface wetting characterization tools previously reported, such as the droplet force apparatus \cite{daniel2017oleoplaning, daniel2019hydration} and the scanning droplet adhesion microscope \cite{liimatainen2017mapping}. However unlike previous approaches, our approach does not require any specialized instrument beyond a conventional AFM setup, which are available to many research groups.

The technique developed here has direct relevance in many applications, e.g. in developing anti-fogging surfaces, in understanding surface condensation processes, and in emulsion science, where the wetting properties are dominated by droplets that are micron or even smaller in size.    

In short, the potential of droplet probe AFM as a surface wetting characterization tool remains largely unexplored and we hope this work will simulate further development of the technique and result in new insights in wetting science.  

%


\providecommand{\noopsort}[1]{}\providecommand{\singleletter}[1]{#1}%

\section*{Acknowledgements}
We thank R. H. A. Ras and S. J. O'Shea for useful discussions. The authors are grateful to the Agency for Science, Technology and Research (A*STAR) for providing financial support under the SERC Career Development Award (grant number A1820g0089) and PHAROS Advanced Surfaces Programme (grant number 1523700101 and 1523700104).

\clearpage
\section*{Materials and Methods}

\textbf{Materials.} IP-Dip photoresist was purchased from Nanoscribe Inc, Germany. Propylene glycol monomethyl ether acetate (PGMEA, $>$ 99.5 $\%$), (3-aminopropyl)triethoxysilane (APTES), perfluorodecyltrichlorosilane (FDTS), glycerine, isopropyl alcohol and ethanol were purchased from Sigma-Aldrich. Chromium (Cr) target (99.994 $\%$) was purchased from Kurt J. Lesker Company, USA. Aluminum (Al) target (99.99 $\%$) was purchased from Zhongnuo Advanced Material (Beijing) Technology Co. Ltd, China. All chemicals were used without further purification, unless otherwise stated. The tipless AFM cantilever was purchased from NanoWorld (Switzerland) and has dimensions of 225 $\times$ 28 $\times$ 1 $\mu$m (length, width, and thickness).   \\  

\textbf{Micropillar fabrication.} The hierarchical micropillar surface is created by direct-write, two-photon lithography using the Nanoscribe$\textregistered$ Photonic Professional instrument (Nanoscribe Inc., Germany). The photoresist used (IP-Dip) is a proprietary negative-tone resin from Nanoscribe that can be used to create submicron feature.  

To improve the adhesion of the resin to the fused silica substrate, we first cleaned the substrate with oxygen plasma for 5 mins at 100 W. The silica substrates were then submerged in 2 $\%$ (v/v) APTES in ethanol for 5 mins, followed by rinsing with a 50 $\%$ (v/v) water/ethanol mixture. Subsequently, the substrates were blown dry with nitrogen gas and dried in an oven at 65$^{\circ}$C.  

The micropillars were designed using computer-aided design (CAD) software SolidWorks$\textregistered$ and their dimensions were defined using the DEScribe software. The structures then written on fused silica substrate by direct laser writing (DLW) using an inverted microscope with an oil-immersion lens (Zeiss Plan Apochromat, 63$\times$, NA 1.4) and a computer controlled piezoelectric stage. The DLW process was performed with an erbium-doped, femtosecond laser source with a centre wavelength of 780 nm, pulse repetition rate of 80 MHz and pulse length of 100 fs. The average laser power was around 40 mW and writing speed was 30 mm/s. After writing, sample substrates were developed in PGMEA, followed by isopropyl alcohol for 30 min each, and then air-dried. \\

\textbf{Surface treatment of micropillar.} 2 nm of Cr film was coated on the substrates, followed by deposition of Al film up to 100 nm with a thermal evaporator system (Syskey, Taiwan) operating under high vacuum of $10^{-6}$--$10^{-7}$ Torr. The micropillar with the Al coating was then boiled in DI water close to 100$^{\circ}$ for 20 minutes. This converts the Al coating into nanostructured boehmite layer. 

We then submerge the surface in a solution of 1 wt$\%$ FS-100, 5 wt$\%$ water in ethanol for 1 h at 70$^{\circ}$C. This fluorinates the boehmite layer and hence renders the micropillar surface superhydrophobic \cite{kim2013hierarchical}. \\ 

\textbf{Droplet probe AFM.} To create the microdroplets, we forced the 40 wt$\%$ glycerine solution through the nozzle of a conventional spray bottle onto a superhydrophobic surface. This generates multiple droplets with diameters between 10--80 $\mu$m. 

To ensure that the microdroplet does not spread on the AFM cantilever, we hydrophobize the surface by vapour-phase silanization with fluorinated silane FDTS. Once the cantilever is fluorinated, we can then pick up a microdroplet of the desired size to perform force spectroscopy measurements on another surface of interest. Note that the microdroplet is attached to cantilever much more strongly than to the superhydrobic surfaces. As described in the main text, there is little or no evaporation of the glycerine droplet.

\end{document}